\documentclass[conference]{IEEEtran}

\usepackage{cite}
\usepackage{graphicx}
\usepackage[cmex10]{amsmath}
\usepackage{algorithmic}
\usepackage[font=footnotesize]{subfig}

\usepackage{url}
\usepackage{fontenc}
\usepackage[utf8]{inputenc}
\usepackage{mathtools}
\usepackage{tabularx}
\begin{document}

\title{Rate Optimal design of a Wireless Backhaul Network using TV White
Space}

\author{\IEEEauthorblockN{Ratnesh Kumbhkar, Muhammad Nazmul Islam, 
Narayan B. Mandayam, Ivan Seskar} \IEEEauthorblockA{WINLAB, Rutgers
University\\ 671 Route 1 South, North Brunswick NJ 08902\\ Email:
\{ratnesh, mnislam, narayan, seskar\}@winlab.rutgers.edu}}
\IEEEoverridecommandlockouts
\IEEEpubid{\makebox[\columnwidth]{978-1-4244-8953-4/11/\$26.00~\copyright~2015 IEEE \hfill} \hspace{\columnsep}\makebox[\columnwidth]{ }} 

\maketitle

\begin{abstract}
The penetration of wireless broadband services in remote areas has primarily been limited due to the lack of economic incentives that service providers encounter in sparsely populated areas. Besides, wireless backhaul links like satellite and microwave are either expensive or require strict line of sight communication making them unattractive. TV white space channels with their desirable radio propagation characteristics can provide an excellent alternative for engineering backhaul networks in areas that lack abundant infrastructure. Specifically, TV white space channels can provide ``free wireless backhaul pipes'' to transport aggregated traffic from broadband sources to fiber access points. In this paper, we investigate the feasibility of multi-hop wireless backhaul in the available white space channels by using noncontiguous Orthogonal Frequency Division Multiple Access (NC-OFDMA) transmissions between fixed backhaul towers. Specifically, we consider joint power control, scheduling and routing strategies to maximize the minimum rate across broadband towers in the network. Depending on the population density and traffic demands of the location under consideration, we discuss the suitable choice of cell size for the backhaul network. Using the example of available TV white space channels in Wichita, Kansas (a small city located in central USA), we provide illustrative numerical examples for designing such wireless backhaul network. 
\end{abstract}

\section{Introduction}
The FCC's (Federal Communications Commission) opening up of TV white spaces for unlicensed use, has led to innovations in cognitive radio technology, spectrum sensing as well as novel proposals for dynamic spectrum access. Over a good part of the last decade, there has been a tremendous amount of research on spectrum policy as well as the theory and practice of cognitive radio networks including dynamic spectrum access (DSA) algorithms, networking protocols and software radio platform development (e.g. \cite{Chap11Mobile, Faul05question, gerami10backhaul, mitola99cognet, haykin05cognetbrain, peha05specshare, ileri08dsamodel, kodialam05capmrmc, kyasanur05capmc, shin06coop, etkin07specshare, clemens05intel, hicks04interf, thomas05cognet, huang05newfront, zhang10bwxchange, NazmulWiOpt, ray06cognet, miljanic07winc2r, warp, ettus, geni}). While recent and prospective policy reforms and the wealth of wireless technology innovations hold great promise for realizing the goals of achieving ubiquitous broadband and continued growth in the wireless sector and services, a significant barrier to broadband in rural areas is the lack of appropriate backhaul solutions. The absence of backhaul stems from a lack of fiber infrastructure necessary to reliably enable backbone connectivity to the global Internet.

While wide area cellular broadband networks have exploded in urban areas with high user densities, it has primarily been due to the ability to engineer smaller coverage areas (i.e., cell-splitting) supported by amply available backhaul infrastructure. The penetration of such broadband networks in rural areas has primarily been limited due to the lack of economic incentives that service providers encounter in sparsely populated areas. The ``economies of scale'' do not hold when sizable infrastructure investments need to be made to support a relatively small customer population in a rural area; a fact that is true even in countries like the U.S. which are thought to have abundant infrastructure. While wired telephone services were historically rolled out in U.S. rural areas in the 20th century due to a federal mandate in exchange for AT\&T’s monopoly on wired telephony, no such mandate exists for wireless broadband services. While optical transport networks and associated cloud/grid computing technologies are evolving rapidly, having ubiquitous deployment of these in rural areas is far from being a reality. Wireless backhaul solutions such as based on satellites or using microwave links are expensive and require strict line of sight requirements making them unattractive and unreliable at times. 

\begin{figure}[!t]
  \centering
    \includegraphics[width=0.45\textwidth]{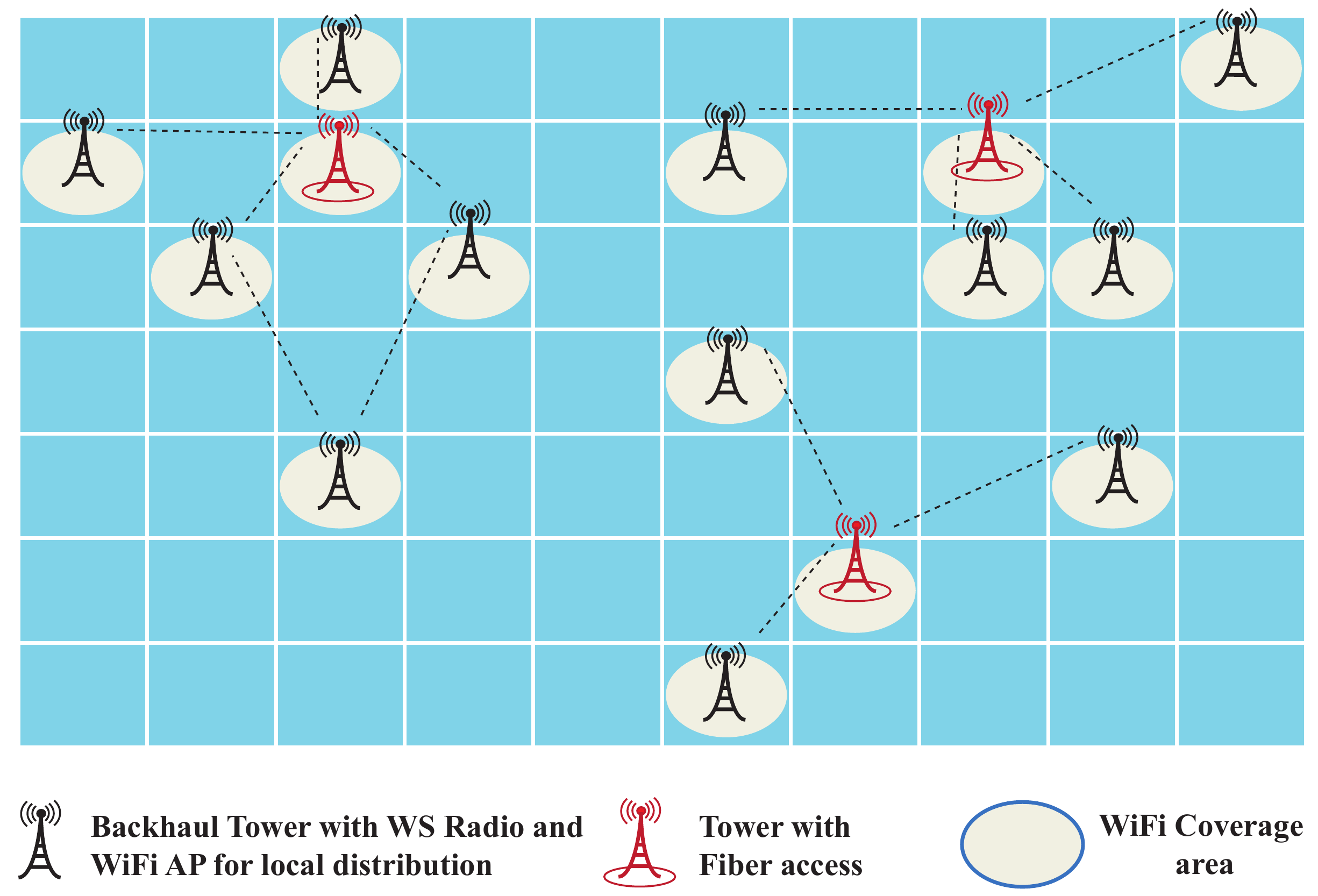}
  \caption{A TV White Space Mesh Network (WSMN) for Rural Backhaul }
  \label{fig:backhaul}
\end{figure}
The scenario we envision is one where local access and distribution to end-users in a rural area will be through WiFi while backhaul enabling Internet connectivity will be through a wide area network of fixed point-to-point TV white space devices on towers \cite{gerami10backhaul} as shown in Figure \ref{fig:backhaul}. Due to the fact that WiFi local distribution will necessarily result in small coverage cells, there needs to be an abundance (spatially) of backhaul infrastructure that can support this system for ubiquitous rural broadband access. We propose to use TV white spaces for the backhaul traffic by designing  a mesh network of white space (WS) radios that can transmit on the available white space channels that are possibly noncontiguous. Further, the uncertainty of available TV white space spectrum poses new challenges in system design along several directions such as planning tower deployments, radio resource management, network management and caching of information.

Our earlier work has motivated the feasibility of rural backhaul using TV white spaces in New Jersey as a case study \cite{gerami10backhaul}.  It is to be noted that when fiber/Ethernet backhaul is mentioned, the usual connotation implies very high bandwidth (e.g. the authors in \cite{dambrosia10100gb} and \cite{wellbrock10100g} discuss 100 Gigabit Ethernet deployments).  We rather envision a scenario where there are areas without other backhaul infrastructure and where a TV white space based network of fixed devices/towers as shown in Figure 1 can serve as a distribution and backhaul network to connect local traffic using various access modalities (e.g. WiFi, and even possibly limited wired or cellular connections) to the Internet (backbone network) at some distance away. The feasibility study in \cite{gerami10backhaul} relies upon several assumptions such as lack of other competing users and interferers in white space as well dynamic changes in traffic, and does not specify any specific modulation, MAC or routing strategies. This paper will investigate the feasibility of multi-hop wireless backhaul in the available white space channels by using noncontiguous Orthogonal Frequency Division Multiple Access (NC-OFDMA) transmissions between fixed backhaul towers. Specifically, we consider joint power control, scheduling and routing strategies to maximize the minimum rate across broadband towers in the network. Depending on the population density and traffic demands of the location under consideration, we discuss the suitable choice of cell size for the backhaul network. Using the example of available TV white space channels in Wichita, Kansas (a small city located in central USA), we provide illustrative numerical examples for designing such wireless backhaul networks. 

\section{Related work}
There exists a large literature in planning towers for wide area wireless systems but these works have been primarily concerned with a specific cellular system and is based on network performance requirements and tower installation costs. For example, \cite{amaldi03umts} focused primarily on supporting UMTS system performance based on signal-to-interference ratio measure, and \cite{yu08plan} considers planning base stations and relay stations in 802.16j networks. NC-OFDMA, an access method with which a sender can transmit using noncontiguous spectrum chunks, has received attention in the literature~\cite{UCSB1, UCSB2,  NCOFDM_Implementation} including our own recent work on routing and scheduling in multihop networks~\cite{Nazmul_NCOFDM1}. However, the issue of spectrum uncertainty due to reuse has not been addressed. Due to the nature of primary and secondary interference in TV white space bands, transmitters will have to use noncontiguous spectrum chunks (or channels) to achieve high data rates. Having multiple radio front-ends to access the noncontiguous chunks can be prohibitive due to cost and real estate available on the transceiver. NC-OFDMA allows nodes to access these noncontiguous spectrum chunks and put null sub-carriers in the remaining chunks \cite{rajbanshi06ncofdmrx}. However, nulling sub-carriers increases the sampling rate (spectrum span) which, in turn, increases the power consumption of radio front ends. Our preliminary work \cite{Nazmul_NCOFDM1} has characterized this trade-off from a cross-layer perspective, specifically by showing how the slope of the ADC/DAC power-bandwidth response influences scheduling and routing decisions in a multi-hop network. In cognitive radio networks research, spectrum reuse (e.g.\cite{liang08sense, srini07cognetdsa}) and routing (e.g.\cite{Cesana11routing, cheng07joint}) algorithms have been proposed to support spectrum coexistence; moreover, there exists a rich history of research in transmit power control for cellular networks, such as \cite{Chiang08powctrl, yates95frameUL, saray02powctrl}. The key differentiators of the work in this paper is the development of reuse and routing algorithms for spectrum coexistence by controlling white space radio channels to provide backhaul for the local traffic demands.  Specifically, we design a rate optimal backhaul network using TV white space channels.

The remainder of the paper is organized as follows: Section \ref{sec:sys} presents the network model. Section \ref{sec:problem} provides the formulation of optimization problem. We present our simulations and results in section \ref{sec:sim}  and we conclude in section \ref{sec:conclude}
\section{System Model}
\label{sec:sys}
\begin{table}[b]
\centering
\caption{List of notations}
\begin{tabular}{|r|p{0.39\textwidth}|}
  \hline
  $\mathcal{N}$ & Set of base stations \\
  \hline
  $\mathcal{A}$ & Set of points with fiber access\\
  \hline
  $\mathcal{E}$ & Set of links \\
  \hline
  $\mathcal{M}$ & Set of total available channel \\
  \hline
  $M$ & Total number of available channels \\
  \hline
  $r_i$ & Data rate at node $i$ due to own traffic \\
  \hline
  $r_{ij}^m$ & Data rate from node $i$ to $j$ on channel $m$\\
  \hline
  $p_{ij}^m$ & Power consumed at node $i$, when transmitting to node $j$ on channel $m$\\
  \hline
  $g_{ij}^m$ & Link gain between node $i$ and node $j$ on channel $m$\\
  \hline
  $x_{ij}^m$ & Scheduling variable for link $ij$ using channel $m$\\
  \hline
  $c_{ij}^m$ & capacity of link $ij$ using channel $m$\\
  \hline
  $s_{ij}^m$ & Signal-to-noise ratio for link $ij$ using channel $m$\\
  \hline
  $N_0$ & Power spectral density of noise\\
  \hline
  $W$ & Bandwidth of each channel\\
  \hline
  $N$ & Number of base stations\\
  \hline
  $A$ & Number of base stations with fiber access\\
  \hline
  $U$ & A large number $\geq$ $P_{max}$\\
  \hline
  $V$ & A large number $\geq$ maximum possible rate\\
  \hline
  $P_I$ & Interference threshold in protocol model\\
  \hline
  $P_{max}$ & Maximum power consumption allowed at each node\\
  \hline
\end{tabular}
\end{table}

\subsection{Network model}
We consider a network comprising of  towers or base stations from  a set $\mathcal{N}$ of size $N$. Each base station is equipped with white space radios which can communicate using TV white space channel. These base stations are placed in a grid structure with $m_1$ rows and $m_2$ columns, i.e. $m_1\times m_2 = N$. Base stations are located at the center of each grid and the length of the sides of each grid   is $l$ km and each base station provides local WiFi access within each grid. Moreover, we do not consider in this paper the explicit design of the WiFi network within each grid. We assume that depending on the size of the grid, WiFi distribution itself can be based on using a mesh network of WiFi nodes. Among these $N$ base stations, a set $\mathcal{A}\subseteq \mathcal{N}$ of size $A$ consists of only those base stations which have access to high speed fiber backhaul links, while remaining base stations in set $\{\mathcal{N}\setminus \mathcal{A}\}$ use only white space channels for backhaul communication. Base stations can use any of the $M$ available white space channels from set $\mathcal{M}$ for the backhaul traffic as long as links satisfy interference constraints. These available white space channels can be noncontiguous and we assume that base stations are capable of accessing these channels using NC-OFDMA. Links are represented as ordered pairs, i.e. link $ij$ represents communication from node $i$ to node $j$, while link $ji$ represents communication from node $j$ to node $i$. We assume a scheduling indicator variable $x_{ij}^m$ such that 
\begin{equation*}
    x_{ij}^m=\begin{cases}
               1, \;\;\;\text{Link $ij$ is scheduled on channel $m\in\mathcal{M}$}\\
               0,\;\;\;\text{Otherwise}
            \end{cases}
\end{equation*}
All communications are assumed to be half duplex and for any channel $m\in \mathcal{M}$, transmission to multiple base stations or reception from multiple base stations is not possible. These constraints can be written as follows
\begin{equation}
  \sum_{j\in \mathcal{N}, j\neq i} x_{ij}^m+\sum_{k\in \mathcal{N}, k\neq i} x_{ki}^m \leq 1 \;\;\; \forall \, i\in \mathcal{N}, \; \forall\, m\in \mathcal{M}.
\end{equation}
\begin{table}[b]
  \centering
  \begin{tabularx}{0.48\textwidth}{|X|X|X|}
    \hline
    Devices & Devices per household, 2012 & Compound Annual Growth Rate, 2012-2017\\    
    \hline
    \hline
    Personal computers & 1.2 & -10\% \\ 
    \hline  
    Smartphones & 1.2 & 20\% \\
    \hline
    Tablets & 0.5 & 40\%\\
    \hline
    Web-enabled TV & 0.5 & 30\%\\
    \hline
    Set Top boxes & 1.0 & 10\%\\
    \hline
  \end{tabularx}
\caption{Parameters used in Cisco VNI predictor}
  \label{table:predict}
\end{table}
The power used for transmission on link $ij$ using channel $m$ is represented by $p_{ij}^m$. The total power used at any base station for transmission on a white space channel is limited by $P_{max}$, therefore
\begin{equation}
  \sum_{j\in \mathcal{N}}\sum_{m \in \mathcal{M}} p_{ij}^m \leq P_{max} \;\; \forall \; i\in \mathcal{N},
\end{equation}
Based on the power allocated for any link $ij$ on channel $m$, this link can support a rate $r_{ij}^m$ which is limited by the capacity constraint as 
\begin{align}
  r_{ij}^m &\leq c_{ij}^m = W \log(1+s_{ij}^m) \;\;\; \forall \, i,j \in \mathcal{N}, \; i\neq j, \\ 
  s_{ij}^m &= \frac{p_{ij}^m g_{ij}^m}{N_0 W}\;\;\; \forall i,j \in \mathcal{N}, \, i \neq j,\nonumber
\end{align}
where $g_{ij}^m$ is the link gain, $N_0$ is the noise  spectral density, and $W$ is the bandwidth of a single channel. The data rate requirement of  the traffic generated within coverage area of a base station $i$ is denoted by $r_i$, therefore according to conservation of flow
\begin{align}
  \sum_{j\in \mathcal{N}}\sum_{m\in \mathcal{M}} r_{ij}^m &= r_i+\sum_{k\in \mathcal{N}}\sum_{m\in \mathcal{M}} r_{ki}^m\;\; \forall \; i\in \{\mathcal{N}\setminus\mathcal{A}\}\\
  \sum_{i\in \{\mathcal{N}\setminus \mathcal{A}\}}r_i &=\sum_{i\in \mathcal{N}} \sum_{j\in \mathcal{A}} \sum_{m\in\mathcal{M}} r_{ij}^m
\end{align}
It can be observed from previous equations that all elements of $\mathcal{A}$ act as sinks in a network, i.e. all traffic generated by towers in $\mathcal{N}\setminus\mathcal{A}$ is equal to traffic entering towers in $\mathcal{A}$. Towers in set $\mathcal{A}$ do not transmit to other nodes using white space channels, therefore the power allocated to these base stations are zero. It can also be noticed that $x_{ij}^m$ is $1$ and $r_{ij}^m$ has non-zero value only when allocated power $p_{ij}^m$ is non-zero. If $U$ and $V$ are sufficiently large numbers then
\begin{equation}
  p_{ij}^m \leq Ux_{ij}^m, \;\;\;r_{ij}^m \leq Vx_{ij}^m,
\end{equation}

To model the interference between different base stations we use a slightly modified protocol interference  model. Under this  model, transmission on link $ij$ using channel $m$ is successful if and only if on the same channel  the interference at base station $j$ from any other transmitting node $k$ is less than some threshold $P_I$. This can be written as 
\begin{equation}
  p_{kl}^m +\left(P_{max}-\frac{P_I}{g_{kj}^m}\right)x_{ij}^m \leq P_{max}\;\;\;\forall \, k\in\mathcal{N},\, l\in\mathcal{N}
\end{equation}
 While calculating the capacity we had ignored the interference caused by other nodes, but this is taken care of by choosing a small enough threshold  $P_I$ for interference in our protocol interference model.

\begin{figure}[!t]
  \centering
\includegraphics[width=0.46\textwidth]{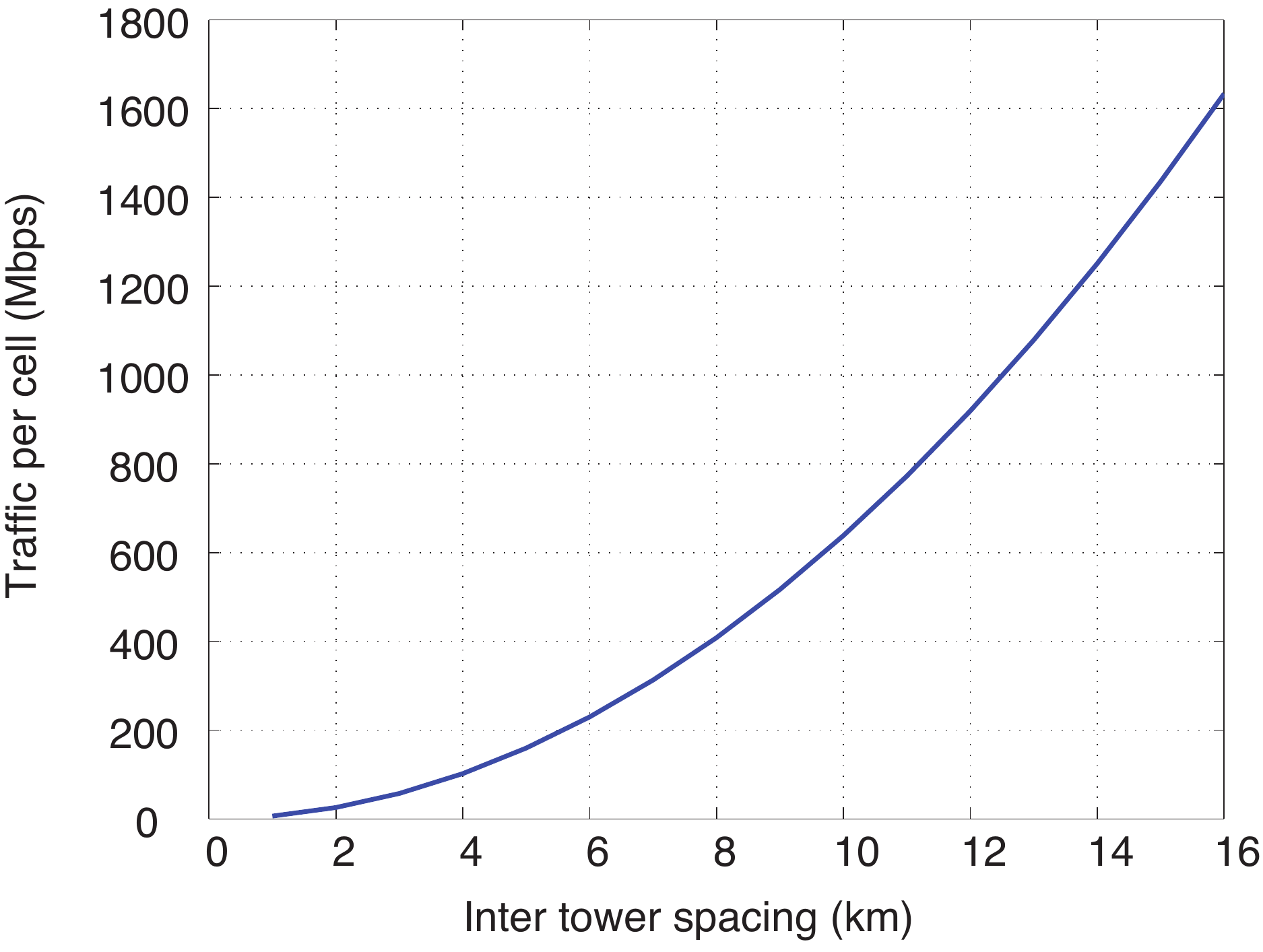}
  \caption{Estimate of  data traffic per cell in year 2017}
  \label{fig:traffic2018}
\end{figure}
The  data rate supported by each base station in its cell should be more than the traffic requirement of that cell's population. To model the traffic demand in a cell, we use the results from a Cisco  study that projects the traffic requirement for 2017. Specifically, the traffic demand is estimated using the  Cisco Visual Networking Index (VNI) \cite{ciscovni} as follows. To predict the rate requirement in 2017, we use  parameters given in Table \ref{table:predict} for  devices in each household for year 2012  and use the projections of growth shown in Table II to arrive at numbers for 2017. Using the default values of application usage (e.g. file sharing, VOIP, gaming, video and data) we obtain an estimate of 130 GB per month per connection (4.33GB  per day). We calculate the traffic demand per cell for different cell sizes assuming penetration of Internet connection  to $78\%$ of total population (using ITU ICT statistics for the developed countries \cite{ituict}). We assume that at any time only $25\%$ of the population is active on the Internet and each household (therefore each subscription) consists of $3$ members. Under the above assumptions, the estimate for per-cell mobile data traffic in year 2017 for different cell sizes is shown in Figure  \ref{fig:traffic2018}. We use this estimate for modeling the traffic demand and find the feasibility of a network using TV white space.
\subsection{Link gain model}
We use the ITU terrain propagation model to calculate the link gain between the base stations for different channels \cite{iturr}. This is a good model to represent  rural areas which might include buildings as well as open fields. Under this model, the median path loss for regular obstructions for LOS communication is given by

\begin{equation}
  PL=FSL+A_d \;\;\;\; dB,
\end{equation}
where the free space path loss is given by
\begin{equation}
  FSL=20\log_{10}d+20\log_{10}1000f+32.44 \;\;\;\; dB
\end{equation}
The distance $d$ between base stations is in km, the frequency $f$ is in GHz. The excess path loss $A_d$ due to diffraction is given as 
\begin{equation}
  A_d=\frac{-20h}{F1}+10\;\;\;\;dB,
\end{equation}
where $h$ is the height difference between the most significant path blockage and the LOS path between the base stations; and $F_1$ is the radius of the first Fresnel zone given by
\begin{equation}
  F_1=17.3\sqrt{\frac{d_1d_2}{fd}}\;\;\;\; m,
\end{equation}
where $d_1$ and $d_2$ are distances of obstruction from base stations in km. The complete link gain between two base stations and resulting SNR are  given by
\begin{align}
  G_{link}&=-PL-N_f+G_{TX}+G_{RX}\;\;\;dB\\
  SNR &= P_T +G_{link}-10\times \log_{10}N_0W\;\;\;dB
\end{align}
where $G_{TX}$ and $G_{RX}$ are the antenna gains  at transmitting and receiving base stations respectively, and $P_T$ is the transmit power.

\section{Problem formulation}
\label{sec:problem}
\begin{figure}[!t]
  \centering
	\includegraphics[width=0.47\textwidth]{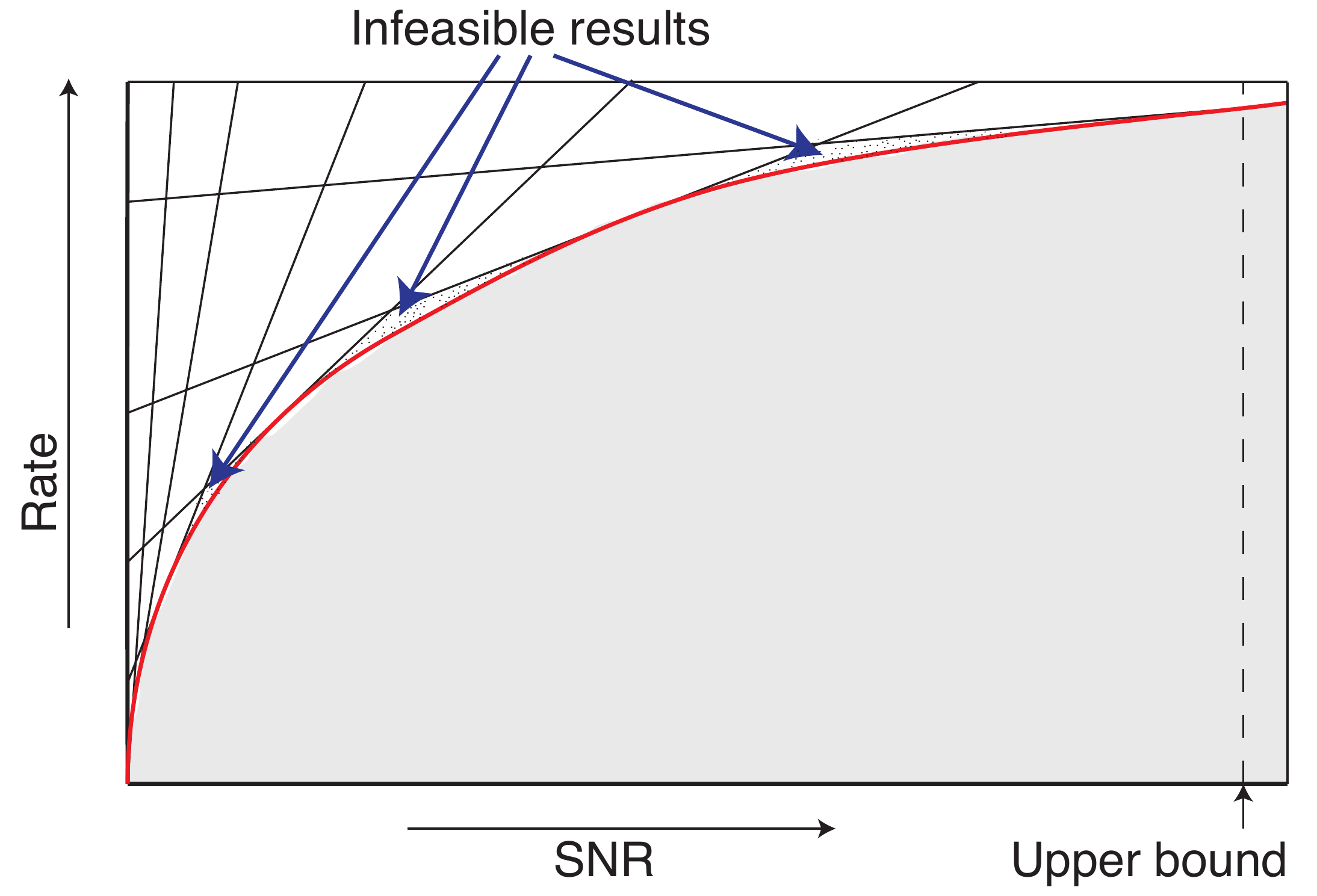}
  \caption{Linear approximation of capacity curve}
  \label{fig:cap}
\end{figure}

In this section we solve the following optimization problem which maximizes the minimum of supported rate $r_i$ at each base station $i$ while performing joint power allocation, scheduling and routing, subject to various power and flow constraints discussed in the previous section.
\begin{equation*}
\begin{aligned}
&{\text{maximize}}\;\;\; \min{r_i} \\
&\text{subject to :}\\
&\sum_{j\in\mathcal{N}}\sum_{m \in \mathcal{M}} p_{ij}^m \leq P_{max} \;\;\forall \; i\in \mathcal{N}\\
&p_{ij}^m = 0 \;\;\; \forall \, i\in\mathcal{A},\, j\in \mathcal{N},\, m\in \mathcal{M}\\
&r_{ij}^m \leq W \log_2(1+s_{ij}^m) \;\;\; \forall \, i,j \in \mathcal{N}, \; i\neq j, \;s_{ij}^m = \frac{p_{ij}^m g_{ij}^m}{N_0 W}\\
&\sum_{j\in\mathcal{N}}\sum_{m\in \mathcal{M}} r_{ij}^m = r_i+\sum_{k\in\mathcal{N}}\sum_{m\in \mathcal{M}} r_{ki}^m\;\; \forall \; i\in\{\mathcal{N}\setminus\mathcal{A}\}\\
&\sum_{j\in \mathcal{N}, j\neq i}x_{ij}^m+\sum_{k\in \mathcal{N}, k\neq i} x_{ki}^m \leq 1 \;\;\;\forall \, i\in \mathcal{N}, \; \forall\, m\in\mathcal{M}\\
&\sum_{i\in \{\mathcal{N}\setminus\mathcal{A}\}}r_i =\sum_{i\in \mathcal{N}} \sum_{j\in \mathcal{A}} \sum_{m\in\mathcal{M}} r_{ij}^m\\
&p_{kl}^m +\left(P_{max}-\frac{P_I}{g_{kj}^m}\right)x_{ij}^m \leq P_{max}\;\;\;\forall \, k,l\in\mathcal{N},\,m\in \mathcal{M}\\\
& p_{ij}^m \leq Ux_{ij}^m \;\;\; \forall \, i\in\mathcal{N},\, j\in \mathcal{N},\, m\in \mathcal{M}\\
& r_{ij}^m \leq Vx_{ij}^m \;\;\; \forall \, i\in\mathcal{N},\, j\in \mathcal{N},\, m\in \mathcal{M}\\
&x_{ij}^m\in \{0,1\}, \; p_{ij}^m \geq 0,\; r_{ij}^m \geq 0,\; \forall \, i,j\in \mathcal{N},\, m\in \mathcal{M}\\
&r_i\geq0 \;\;\;\forall i\in \mathcal{N}
\end{aligned}
\end{equation*}
This is a mixed integer nonlinear program, since $x_{ij}^m$ variables are binary and capacity constraints are nonlinear. We modify the nonlinear capacity constraint to multiple linear constraints as piecewise linear approximation. To get this approximation, we use  values of SNR based on $P_{max}$  as an upper bound.  Starting with 0 dB SNR, each  $\sim 10$ dB increment is approximated by a line, which is tangential to the capacity curve as shown in Figure  \ref{fig:cap}. 

We solve this modified mixed-integer linear program  using MOSEK with CVX \cite{cvx, mosek}. MOSEK solves the mixed-integer program using the branch-and-bound method, which has exponential complexity. As seen in Figure  \ref{fig:cap}, MOSEK might result in an infeasible solution due to linearization of  the capacity constraint. If an infeasible solution occurs, then corresponding SNR is calculated using power variables $p_{ij}^m$ and then the new rate is calculated with this SNR as
\begin{equation}
  r_{ij,\text{new}}^m = W\log_2\left(1+\frac{p_{ij}^mg_{ij}^m}{N_0W}\right)
\end{equation}
The supported data rate $r_i$ at the base station $i$ is  decreased by $r_{ij}^m-r_{ij,\text{new}}^m$ to maintain the flow conservation, and after checking the feasibility of all rate variables a new minimum supported rate is calculated.

\section{Simulation and results}
\label{sec:sim}
\begin{figure}[!t]
  \centering
\includegraphics[width=0.35\textwidth]{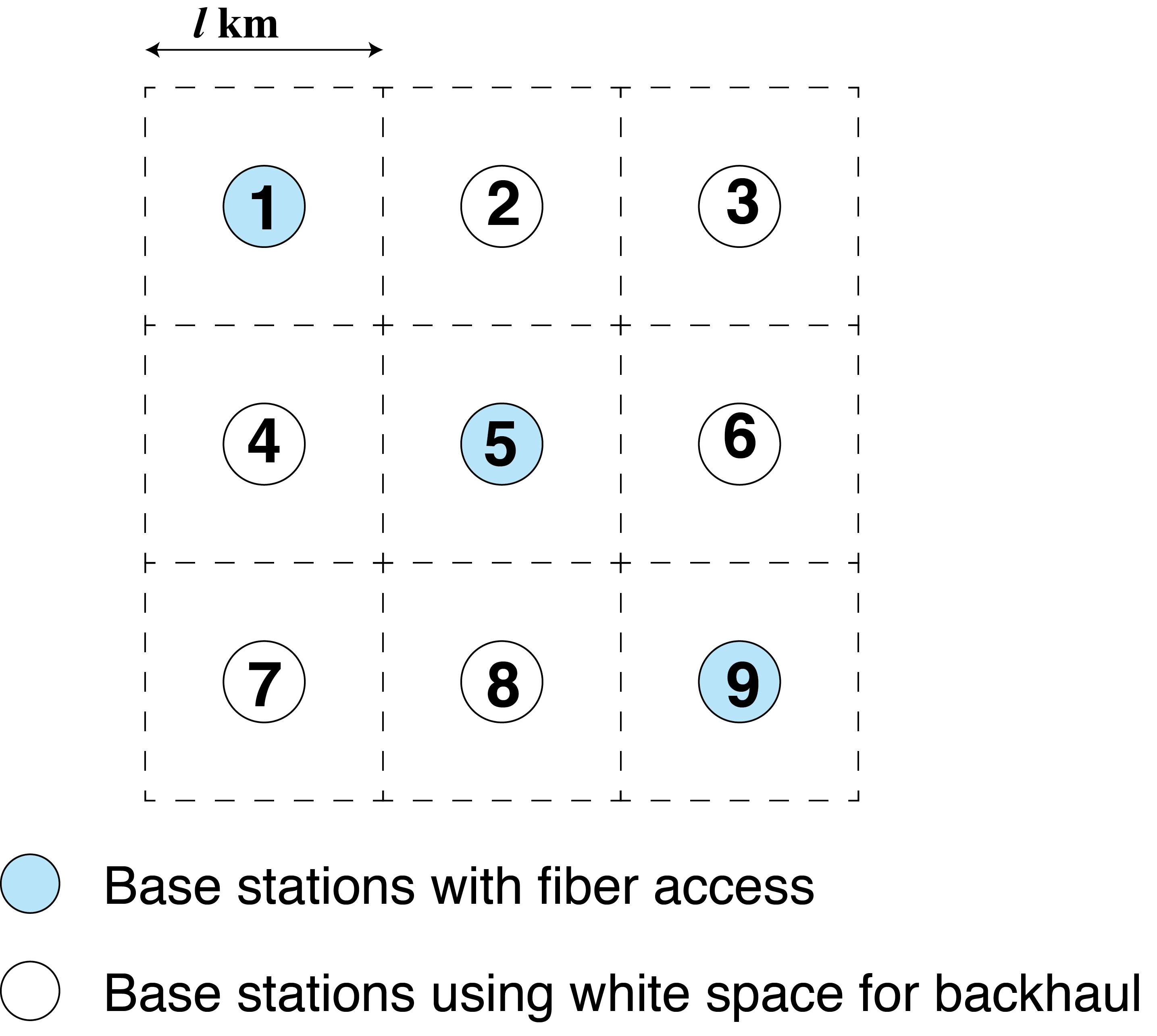}
  \caption{Simulation setup with $\mathcal{A}=\{1,5,9\}$}
  \label{fig:grid}
\end{figure}

We consider a small network of 9 base stations located in a $3\times 3$ grid as shown in Figure  \ref{fig:grid}, covering total area of $9l^2$ sq. km in our simulation.  Figure \ref{fig:grid}  shows base stations 1, 5 and 9 with fiber access as an example, while the other base stations use only white space channels  to backhaul their traffic to one of the fiber access destinations. Table \ref{table:simpar} shows the values of different parameters used in our simulation. There are seven white space channels available for fixed devices in Wichita, KS, USA, which we use in our simulation
\begin{equation*}
\mathcal{M}=\{57,79,85,491,527,533,671\}\;\;\text{MHz}
\end{equation*}
We vary the grid size  and number of towers with fiber access points to check the feasibility of white space based backhaul. Different  values of $l$
are chosen from a set $\mathcal{L}$, where
\begin{equation*}
\mathcal{L}=\{2,3,4,5\}\;\;km
\end{equation*}
thus covering areas of $36$, $81$, $144$ and $225$ sq. km respectively, while  values of $\mathcal{A}$  are chosen to be any of the following sets
\begin{align*}
  \mathcal{A}_1 &= \{1,5,9\},\\
  \mathcal{A}_2 &= \{1,3,7,9\},\\
  \mathcal{A}_3 &= \{1,3,5,7,9\}
\end{align*}
\begin{table}[t]
  \centering
  \begin{tabular}{r|l}
\hline
    \hline
    $P_{max}$ & 4 watts \\ 
    \hline
    Noise figure $N_f$ & 10 dB\\
    \hline
    $d_1$ & d/2\\
    \hline
$d_2$ & d/2\\
    \hline
    $G_{TX}$ & 6 dB\\
    \hline
    $G_{RX}$& 6 dB\\
    \hline
    Height of link & 30m\\
    \hline
    Height of obstruction & 15m\\
    \hline
    Channel bandwidth $W$ & 6 MHz\\
   \hline
  \end{tabular}
\caption{Parameters used in simulation}
  \label{table:simpar}
\end{table}
Due to the complexity of the mixed-integer linear program we use an approximate version of this problem, which gives the highest feasible value of $r_i$ with a granularity of 1Mbps. We use the traffic demand model described earlier to generate traffic for the entire 3x3 grid and check the feasibility of using white space based backhaul with different grid sizes and different values of $\mathcal{A}$.

\begin{figure}[!t]
  \centering
\includegraphics[width=0.35\textwidth]{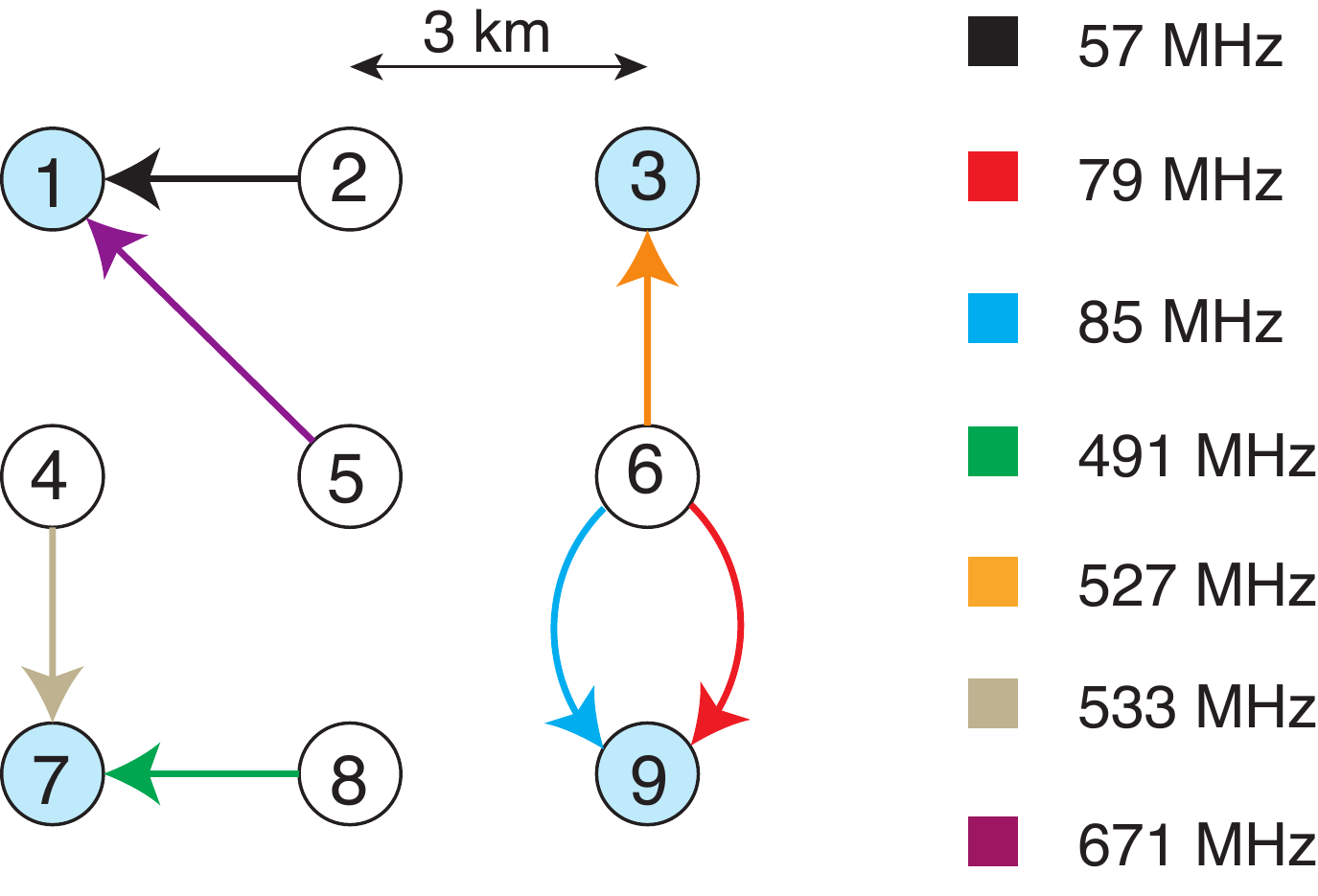}
  \caption{Routing and scheduling obtained  for  $\mathcal{A}=\{1,3,7,9\}$ with $l=3$km}
  \label{fig:simgrid}
\end{figure}

Figure \ref{fig:simgrid} shows a frequency scheduling and routing scheme that results from solving the optimization problem which achieves a minimum supporting data rate of 75 Mbps at each base station, where the grid size is $3 km$$\times$$3 km$ and $\mathcal{A}=\{1,3,7,9\}$. 

\begin{figure}[!t]
  \centering
\includegraphics[width=0.45\textwidth]{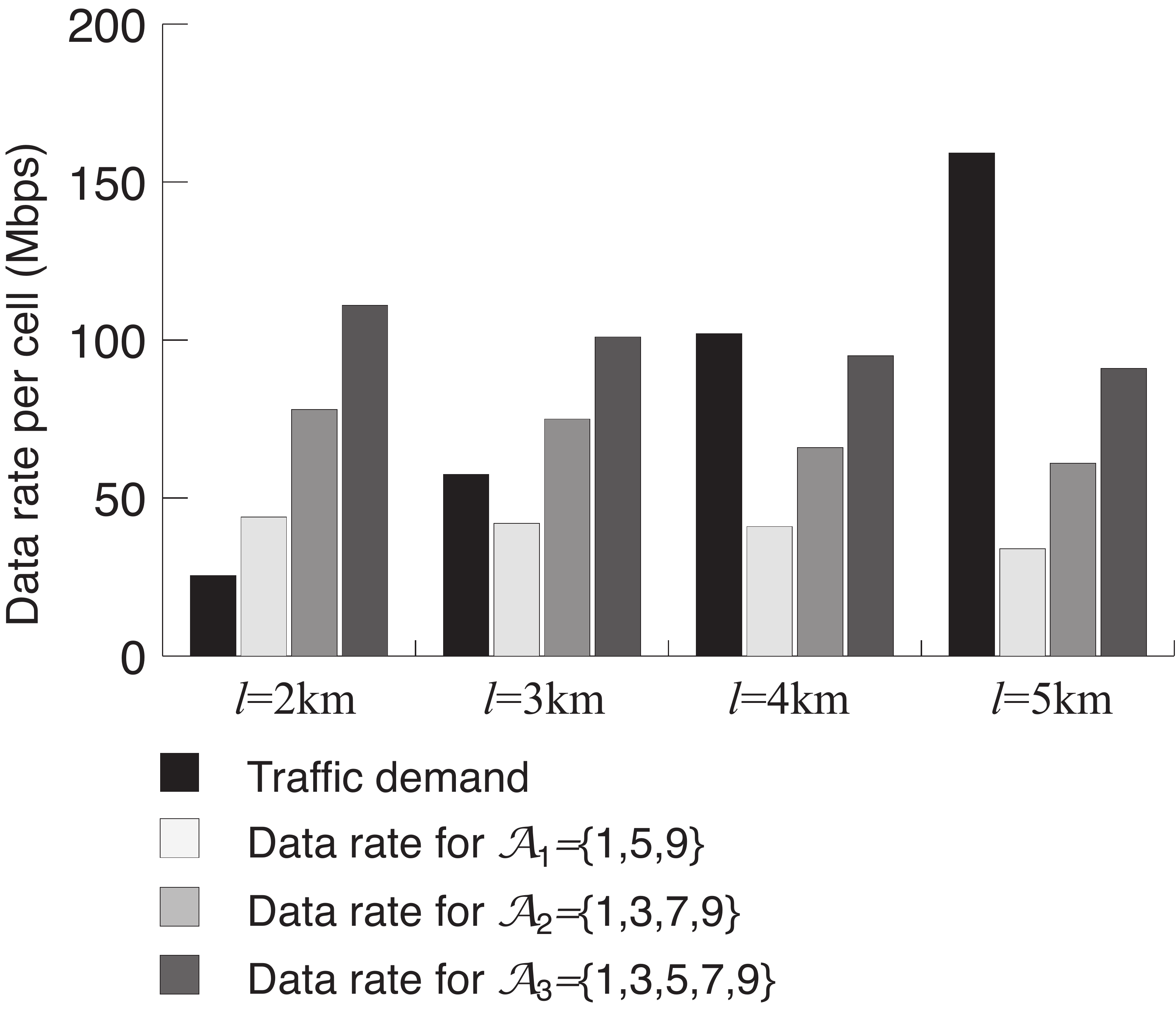}
  \caption{Data rate supported at each cell for different grid size}
  \label{fig:resbar}
\end{figure}

Figure \ref{fig:resbar} shows the complete result obtained from solving the optimization problem for different values of the grid size and different values of set $\mathcal{A}$. The data traffic demand per cell increases as the cell size increases  due to increase in number of users per cell. It can be seen that for $l=2km$, use of $\mathcal{A}_1$, $\mathcal{A}_2$ or $\mathcal{A}_3$ as fiber access points meets the traffic demand. However, as the cell size increases, the traffic requirement is met only when the number of fiber access points is large. For $l=5km$ and $l=4km$, the traffic demand is not met by any of the fiber access configurations used in the simulations, but $l=3km$ presents an interesting result. For $l=3$km, the traffic demand is met by using  $\mathcal{A}=\mathcal{A}_2$ or $\mathcal{A}=\mathcal{A}_3$. Note that the case of $\mathcal{A}=\mathcal{A}_2$ represents a scenario where only 4 base stations with fiber access (out of a total of 9) can provide backhaul for a coverage area of 81 sq. km.

\section{Conclusion} 
\label{sec:conclude}
The penetration of wireless broadband services in remote areas has
primarily been limited due to the lack of economic incentives that
service providers encounter in sparsely populated areas. Besides,
wireless backhaul links like satellite and microwave are either
expensive or require strict line of sight communication making them
unattractive. TV white space channels with their desirable radio
propagation characteristics can provide an excellent alternative for
engineering backhaul networks in areas that lack abundant
infrastructure. Specifically, TV white space channels can provide ``free
wireless backhaul pipes'' to transport aggregated traffic from broadband
sources to fiber access points. This paper investigated the feasibility of multi-hop wireless backhaul using the available TV white space channels in Wichita, Kansas (a small city in central USA) as an example.  We performed joint power control, scheduling and routing to maximize the minimum rate across broadband towers in the network. Simulation results showed that a service provider could bring fiber to a subset of deployed towers but still could meet the backhaul requirement of each tower using white space channels.

Our work can be extended to minimize the fiber layout cost while supporting backhaul requirement of each tower in a rural area using TV white space channels. Since noncontiguous spectrum access increases the power consumed in the ADC and DAC transceiver circuitry due to increased spectrum span, it is also of interest to study the power consumption in such backhaul networks using noncontiguous spectrum access. Further the impact of interference from other unlicensed devices and its implication for backhaul capacity is also of interest.


\bibliographystyle{IEEEtran}
\bibliography{comsnets} \end{document}